\documentstyle[editedvolume,numreferences,epsfig
]{crckapb}

\begin{opening}
\title{Transport through a single--band wire connected to
measuring leads}
\author{In\`es Safi}
\author{H.J. Schulz}
\institute{Laboratoire de Physique des Solides\\
Universit\'e Paris--Sud, 91405 Orsay, France}
\end{opening}
\runningtitle{wire connected to leads}

\begin{document}

\begin{abstract}
{Transport through a one-dimensional wire of interacting electrons connected to semi infinite leads is investigated using a bosonization approach. The dynamic nonlocal conductivity is rigorously expressed in terms of the transmission. For abrupt variations of the interaction parameters at the junctions, an incident electron is transmitted as a sequence of partial charges: the central wire acts as a Fabry-P\'erot resonator. The dc conductance is shown to be given by the total transmission which turns out to be perfect. When the wire has a tendency towards superconducting order, partial Andreev reflection of an incident electron occurs. Finally, we study the role of a weak barrier at one contact or inside the wire by a renormalization group method at finite temperature. We compute the conductance in the presence of localized or extended disorder. The results are compared to recent experiments.}
\end{abstract}

\section{Introduction}

Quantum wires, obtained by lateral confinement of a two dimensional electron
gas \cite{kastner}, provide a good candidate to test the well-developed
theory of one-dimensional interacting systems which show the so-called
Luttinger liquid behavior
\cite{haldane_bosonisation,schulz_revue}. According to \cite{apel,kane_fisher} the interactions should renormalize the conductance
$g$ of a pure wire: $ g=2g_0K$ where $g_0=e^2/h$ is the conductance quantum
and $K$ is a key parameter depending on interactions, with $K=1$ for a
non--interacting system. The reduction of the conductance in the presence of
localized or extended disorder has a power-law dependence on temperature
\cite{lut_bos,apel_rice,giamarchi_loc}, or on the wire length \cite{ogata},
determined also by $K$. Recent experiments \cite{tarucha} on wires with
a length up to $ 10\mu m$ show indeed a power law, and the authors extract $
K=0.7$. At high temperature, where the effect of impurities is less
pronounced, the conductance gets very close to the ballistic value $2g_0$,
i.e. it is different from the expected $2K g_0 = 1.4 g_0$. The latter result
is obtained by computing the current as a response to an electric field
restricted to a finite segment of an infinite interacting wire \cite{kane_fisher,schulz_revue}. However, in a mesoscopic device the
boundaries intervene in a nontrivial way: one then has to change some
standard concepts on resistance of macroscopic systems.

An intuitive approach to transport was pioneered by
Landauer \cite{landauer:70} and developed by B\"{u}ttiker \cite{butticker_revue} and others: the
conductance of a coherent sample ought to be proportional to the ease with
which electrons can be transmitted through it:
\begin{equation}
g=g_0T ,  \label{land}
\end{equation}
where $T$ is the transmission coefficient. Then the finite resistance of a
perfect system can be traced to its interface with the reservoirs
\cite{imry}. Thus it is necessary to view the system formed by a mesoscopic
sample and its contacts as a whole \cite{butticker_d.c.}. Furthermore,
B\"{u}ttiker \cite{butticker_pretre} stresses the importance of the inclusion
of the nearby metallic bodies in any discussion of the
a.c. conductance. When attempting to derive a Landauer formula rigorously,
one has to face the problem of modeling the reservoirs properly, a point
which has lead to different points of view in the literature. Many authors
 \cite{fisher_lee,stone_houches} include perfect leads between the sample and
the reservoirs: this allows to define the incoming and outgoing scattering
states, thus the transmission coefficients. However, this description becomes difficult if the electrons
interact in the sample. The extension of Landauer's formula to interacting
electrons is mostly phenomenological \cite{butticker_resistors}. Apel \cite{apel} derived it if disorder is included on a finite segment of an
infinite interacting one-dimensional wire, but this is not necessarily a
realistic situation. For instance, a quantum wire typically opens smoothly
into two wide leads formed by the same two-dimensional gas. While the
electronic correlations are enhanced in the confined region, they may well
be negligible outside of it.

\begin{figure}[htb]
\begin{center}

\epsfig{file=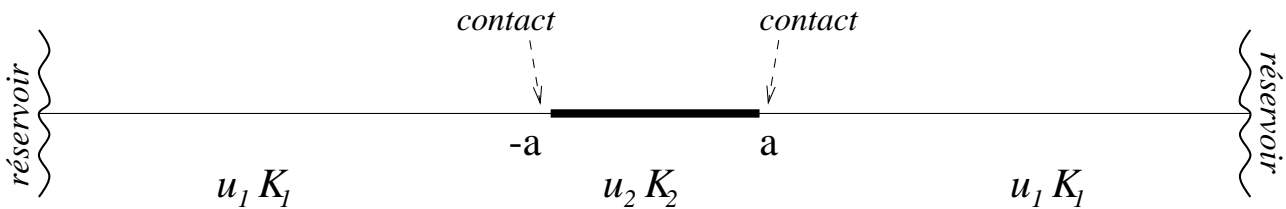,width=17 cm}

\caption{An interacting one-channel wire connected perfectly to very long leads}\label{fil}
\end{center}\end{figure}

It seems difficult to treat exactly interactions if one includes the
complications of the experimental setup. Thus we suppose that the wire is
connected to reservoirs via very long perfect one-dimensional leads where
the electrons do not interact (see fig.(\ref{fil})). Neglecting the interactions in the leads might
be appropriate as a primitive model of Fermi liquid behavior in the
two--dimensional gas forming the contacts. The effect of the reservoirs is
accounted for only through the voltage they impose in a conductance
measurement, and we assume this voltage not to be affected by the
current. The infinitely long leads insure that a transmitted electron never
returns to the wire and therefore cannot interfere with either the wire or
the incident current. Even though this is an oversimplification, it permits
us to treat the wire and the leads on the same footing, and to treat exactly
interactions in the wire using the bosonization method. In our model the
contacts are perfect, and delimit the wire where short-range
electron--electron interactions exist. Note that our approach is opposite to
that of Fabrizio and Gogolin \cite{fabrizio_open} who consider an isolated
interacting wire, and disturb it by a weak tunneling into the ends. We
investigate the effect of a weak backscattering potential at one contact, as
well as an extended disorder. The extension of Landauer's
formula to the interacting dirty wire will be presented. Finally, we will see how our
space-dependent interactions model can give rise to phenomena familiar from
metal-superconductor interfaces.\\
\\
\section{The Model}

For
simplicity, we consider only spinless electrons with short-range
interactions, which are described by the Hamiltonian \cite{haldane_bosonisation}
\begin{equation}
H=\int_{-L}^L\frac{dx}{2\pi }\left[ uK(\partial _x\Theta )^2+\frac
uK(\partial _x\Phi )^2\right]  \label{H}
\end{equation}
where the boson field $\Phi $ is related to the particle density through: $
\rho -\rho _0=-\partial _x\Phi /\pi $, and $\partial _x\Theta /\pi $ is the
conjugate momentum field to $\Phi $: $\left[ \partial _x\Theta(x) ,\Phi
(y)\right] =i\pi \delta (x-y)$. There are no single-particle
excitations, but sound wave excitations with velocity $u$. Some intuition of
the Luttinger liquid features can be
gained by noting that an injected right-going electron decomposes into a
charge $(1+K)/2$ propagating at velocity $u$ and a charge $(1-K)/2$
propagating at $-u$. The total resulting current is $K$.

We now consider a finite interacting wire perfectly connected to two
identical leads at its end points $\pm a$ (see fig.(\ref{fil})). We shall label the quantities
pertaining to the leads (central wire) by the subscript $1$ ($2$). In $H$
(eq.(\ref{H})) the parameters $u,K$ then vary from $u_2,K_2$ in the wire to
$ u_1,K_1$ outside: most of our results will concern abrupt variations at
$\pm a$, and the physically most relevant situation of noninteracting leads:
$K_1=1$. We adopt periodic boundary conditions on the whole system. This is a
mathematical convenience. At scales much less than the total length $2L$,
the physics in the intermediate region does not depend on the boundary
conditions at the end points of the leads, and is expected to be the same as
for an open system. We'll henceforth discuss the results once $L$ has been
taken larger than any other length.

\section{The transmission process}

\begin{figure}[htb]
\begin{center}
\mbox{\epsfig{file=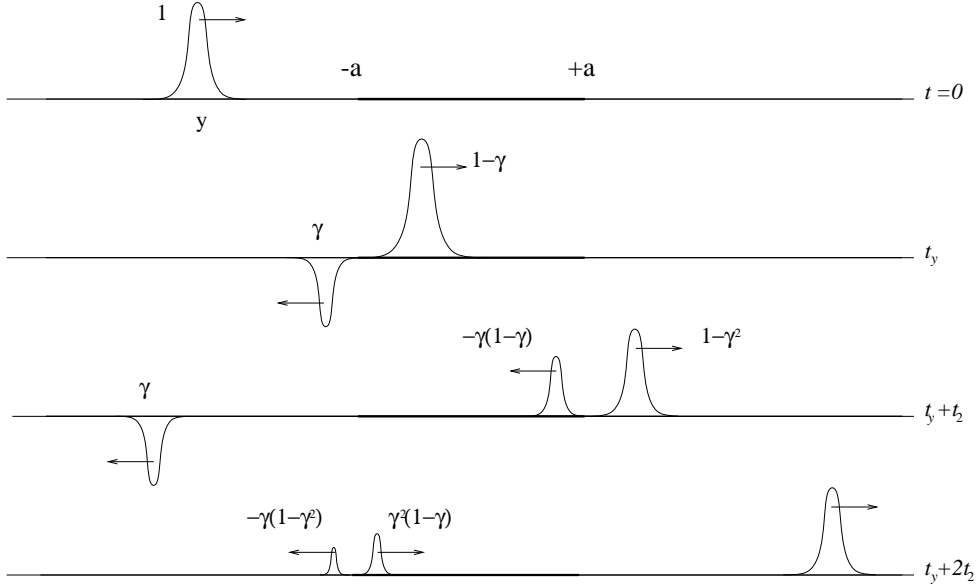,width=13 cm}}
\caption{The transmission process of an incident electron on the wire in the
case where $K_2>1$ and $u_1=u_2$. We denote: $t_y=(-a-y)/u_1$. At $t_y+(2n+1)t_2$ (resp. $t_y+2nt_2$), a charge
$\gamma^{2n}(1-\gamma^2)$(resp. $-\gamma^{2n-1}(1-\gamma^2)$) comes out $a$(resp.$-a$). The first
reflected charge is of hole type, while the subsequent ones are of electron
type. If $K_2<1$, the hole and electron type reflected charges are exchanged.}
\end{center}\label{transmission}
\end{figure}

We consider a thought experiment where one injects a right-moving electron
on the left perfect lead (at a point $y\leq -a$) and places a detector on
the right lead (at $x\geq a$) which measures the transmitted charge $
M_{++}(x,y,t)$ at time $t$. According to the relation $\rho -\rho
_0=-\partial _x\Phi /\pi $, creating an electron amounts to induce a kink in
$\Phi $: the propagation of our electron is determined once the equation of
motion for $\Phi $ is solved. Using the Hamiltonian (\ref{H}), this equation
is
\begin{equation}
\partial _{tt}\Phi -uK\partial _x\left(\frac{u}{K}\partial _x\Phi \right)=0   \;.
\label{motion}
\end{equation}
The inverse of the differential operator acting on $\Phi $ is nothing but
the correlation function $G(x,x^{\prime },t)=i\theta (t)\left\langle \left[
\Phi (x,t),\Phi (x^{\prime },0)\right] \right\rangle $. Its knowledge allows
us to determine the time evolution of the operators of interest for us:
\begin{equation}\label{jr}
\widetilde{\rho }_{\pm }=\frac 12(\rho \pm j/u)=\frac 1{2\pi }(-\partial
_x\pm \frac 1u\partial _t)\Phi
\end{equation}
In the noninteracting leads, $\widetilde{\rho }_{\pm }$  are simply the right and
left-going electron density. Given the initial conditions $\left\langle
\widetilde{\rho }_{+}(x,0)\right\rangle =\delta (x-y)$, $\left\langle
\widetilde{\rho }_{-}(x,0)\right\rangle =0$, the transmitted and reflected
charge, $ \left\langle \widetilde{\rho }_{+}(x\geq a,t)\right\rangle
=M_{++}(x,y,t)$ and $\left\langle \widetilde{\rho }_{-}(y^{\prime }\leq
-a,t)\right\rangle =M_{-+}(y^{\prime },y,t)$, can be deduced from $G$.  We
can however predict $ M_{++}$ and $M_{-+}$ by simple arguments if $u$ and
$K$ are step functions. In this case, eq.(\ref{motion}) reduces to a wave
equation with discontinuous velocity. One can convince oneself that
eq.(\ref{motion}) can't lead to a discontinuity neither in $\Phi $ nor in
$(u/K)\partial _x\Phi $. Thus we require the continuity of the current
$j=\partial_t\Phi /\pi $ at the contacts $\pm a$, which is physically
plausible, and the continuity of $(u/K)\rho $. Using the relation
(\ref{jr}), this tells us how our electron will evolve: once it impinges on
the contact at $-a$, it gets reflected with coefficient $\gamma
=(1-K_2)/(1+K_2)$. The transmitted charge, which is a peak in $
\left\langle \widetilde{\rho }_{+}\right\rangle$, will take a time $
t_2=2a/u_2$ until it reaches $+a$, where a charge $1-\gamma ^2$ is
transmitted, while the reflected part will continue to bounce back and forth
so that after each period $t_2$ a partial charge leaves the wire. This
process is illustrated in the figure (\ref{transmission})
. {\em Thus the incident electron is decomposed into a
sequence of spatially distributed charges.} If we wait a long time compared
to $t_2$, the series sums up to unity: $\sum_n\gamma ^{2n}(1-\gamma ^2)=1$.
{\em The total transmission is perfect.}

This argument is confirmed by an exact computation of $G$ for any spatial
arguments. The expression for $M_{++}(x,y,t)$ is given in \cite{ines}. We
give here only its Fourier transform:
\begin{equation}  \label{resonant}
M_{++}(x,y,\overline{\omega })=\exp i\overline{\omega }(t_x+t_y)\frac{
1-\gamma ^2}{\exp (-i\overline{\omega }t_2)-\gamma ^2\exp (i \overline{
\omega }t_2)}
\end{equation}
where $t_x$ is the time it takes for an electron to go from $x$ on the lead
to the closest contact, i.e, $u_1t_x=\left| x\right| -a$. Recall that $
t_2=2a/u_2$ is the traversal time of the central wire. We abbreviate
$\overline{ \omega }=\omega +i\delta $, where $\delta $ ensures the
convergence in the thermodynamic limit: $e^{-\delta L}\ll 1$. It's worth
noting that $ M_{++}(-a,a,\overline{\omega})$ coincides with the total
 transmission of a double scatterer obtained through the composition:

\[\frac{K}{K_1}
\left( 
\begin{array}{cc}
1 & \gamma \\ 
\gamma & 1
\end{array}
\right)
\left( 
\begin{array}{cc}
e^{i\overline{\omega} t_2} & 0 \\ 
0 & e^{-i\overline{\omega} t_2}
\end{array}
\right) 
 \frac{K}{K_2}
\left( 
\begin{array}{cc}
1 &-\gamma \\ 
-\gamma &1 
\end{array}
\right) \]
where $K=(K_1+K_2)/2$. The first matrix is the transfer matrix at $-a$,
the third its inverse at $+a$, between which the propagation matrix is sandwiched. These matrices are acting on $(j_{+}=u\widetilde{\rho }
_{+},j_{-}=-u\widetilde{\rho }_{-})$ and not on the wave function amplitudes
as one is used to. Therefore
each column of a transfer matrix sums up to unity. 
The perfect transmission also appears in eq.(\ref{resonant}) which gives
$M_{++}(x,y,\overline{\omega }\ll \omega _2)\simeq 1$. Accordingly, the
expression for the reflected charges gives $M_{-+}(y^{\prime },y,
\overline{\omega }\ll \omega _2)\simeq 0$. Note also that $
M_{++}(-a,a,n\omega _2)=1.$ The central wire acts as a Fabry-P\'erot
resonator with symmetric mirrors\cite{ines}.
 But the perfect transmission is not
specific to the symmetric structure. We can show that for any internal shape
of $u$ and $K$ within $\left[ -a,a\right] $, varying smoothly or not, the
total transmission is still perfect. This result holds even if we include
interactions on the leads, i.e. for $K_1\neq 1$. In this case, an incident or
transmitted flux is recognized by a peak in $\left\langle \widetilde{\rho
}_{+}\right\rangle $, even if this does not correspond anymore to the
original right-going electrons. Apart from an exact evaluation of $\lim_{\omega \rightarrow
0}M_{++}(x,y,\omega)=1$ and $\lim_{\omega \rightarrow
0}M_{-+}(y^{\prime },y,\omega)= 0$,
the following observation explains the perfect transmission: in the
zero-frequency limit, not only the current $j$ but also $(u/K)\rho $ (see
eq.(\ref{motion})) are uniform along the system. We deduce that
$\left\langle \widetilde{\rho }_{+}(\omega =0)\right\rangle $, as well as
$\left\langle \widetilde{\rho }_{-}(\omega =0)\right\rangle $ take the same
uniform values on the opposite leads since the latter have the same
$u/K$. In case an electron emerges initially from $y\leq -a$, $ \left\langle
\widetilde{\rho }_{+}(x\leq -a,\omega =0)\right\rangle =\int_0^\infty \delta
\left[ (x-y)/u_1+t\right] dt=1\ $ while $\left\langle \widetilde{\rho
}_{-}(x\geq a,t\right\rangle =0$ (there are no electrons coming from the
right lead). Thus the total density of transmitted (reflected) charge is
unity (zero).

\section{Transport in the pure wire}

If we apply a time-dependent potential, the Hamiltonian acquires the
supplementary term $-\int V(x,t)\partial _x\Phi /\pi $. The equation (\ref
{motion}) for $\Phi $ has now a source term on the right hand side,
$E(x,t)$. Thus we get $\langle \Phi (x,t)\rangle=\int \int G(x,y,t-t^{\prime
})E(y,t^{\prime })$; differentiating this relation with respect to time, we
find that the current is exactly linear in the electric field\cite{ines}, and that
$\sigma (x,y,t)=-  g_0\partial _tG(x,y,t)/\pi$. This shows how
transport is related to propagation \cite{ines}. For $x,y$
on opposite leads, the last relation reads
\begin{equation}
\sigma (x,y,t)=g_0M_{++}(x,y,t)  \label{Landidentity}
\end{equation}
This is a generalization of the Landauer formula, eq.(\ref{land}), to a
dynamic situation. When $u$ and $K$ are step functions, $\sigma
(x,y,\overline{\omega })$ can be computed for any spatial
arguments. Its expression is reported in \cite{ines} for points inside the
wire, while it is given by (\ref{resonant}) for points on opposite leads,
owing to the identity (\ref{Landidentity}). 
  
In order to exploit the identity (\ref{Landidentity}) further, we make a
digression. Suppose that the reservoirs set up a uniform oscillating potential 
$V_R(\omega)$ and $V_L(\omega)$ on the right and left leads. Since there are
no scatters inside the wire that can generate an electric field, we expect the potential
drop to be concentrated at the
contacts. Taking the potential on the wire as a reference, the resulting current at any point on
the leads is:
\begin{equation}
j(x,\omega)=\sigma (x,-a,\omega)V_L(\omega)-\sigma(x,a,\omega
)V_R(\omega )  \label{jLR}
\end{equation}
Now let $\delta\ll\omega\ll \omega_2$: this is already a large
frequency domain, of the order of some GHz, for wires with a length of some
$\mu$m. Then the current on the leads oscillates with a spatial period much
longer than the wire length, and we can take it as uniform on each lead (if
measured at a reasonable distance from the contacts). It's 
tempting to interpret eq.(\ref{jLR}), taken for instance at $-a$ and $a$, as the
relations that define the two conductance coefficients in a two-probe
measurement  \cite{butticker_pretre}: $$ g_{LL}(\omega )=g_{RR}(\omega
)=-g_0[1+M_{-+}(-a,-a,\omega )]\simeq-g_0[1+i\omega t_2(K_2^{-1}-K_2)/2]$$  $$g_{RL}(\omega
)=g_0M_{++}(-a,a,\omega) \simeq g_0\left[1+i\omega t_2(K_2^{-1}+K_2)/2\right].$$ However, all this has to be taken with
caution. The voltage profile cannot be known a priori. The interesting
issues in a.c. transport developed by B\"utticker  \cite{butticker_pretre}
cannot be addressed here due to the lack of long-range Coulomb
interactions.
Things are less ambiguous in the zero-frequency limit, where we do not need
to know the exact potential distribution within the wire (if we consider the
total voltage drop as fixed by the reservoirs). This is because $
\lim_{\omega t_2 \rightarrow 0}\sigma (x,y,\omega )=\sigma $ is independent of
its spatial arguments, and thus yields the conductance $g=\sigma $ as may be
checked from: $j(x)=\int \sigma (x,y)E(y)$. This relation shows also that
the uniformity of $\sigma (x,y)$ is a constraint that ensures the uniformity
of the current in a one-dimensional system. We can verify this explicitly in
our model. It turns out that $g=\sigma =g_0$. This result appears from
(\ref{Landidentity}) which becomes a trivial Landauer identity in view
of the perfect transmission (eq.(\ref{land}) with $T=1$). When the leads are
interacting, $g=K_1g_0$: the conductance is renormalized by $K_1$, not $K_2$
as predicted by Kane and Fisher in a different geometry. These authors
pointed out that their results are valid at temperatures or frequencies greater than
$\omega _2$ so as to ensure that one is not measuring the properties of the
external leads. At first sight, this condition does not come out of our
analysis: all the previous results are independent of temperature, and there
are oscillations at $\omega >\omega_2$. But our implicit assumption was that $\delta ^{-1}$,
the adiabatic turn-on time of the external field, is greater than the
traversal time $ t_2=2\pi /\omega _2$. It's thus interesting to see what
we get in the opposite limit $\delta t_2\gg 1$. Then the expression of the
conductivity \cite{ines} has different zero-frequency
limits depending on its spatial arguments (the steady current gets
homogenized only on scales of the order $\delta ^{-1}\ll t_2$). Near the
center of the wire, i.e. for $|x|\ll u_2/\delta$, we have $\sigma
(x,y)=g_0K_2$, while near the contacts, i.e. $|x\pm a| \ll u_2/\delta$,
$\sigma (x,y)=g_0K_{a}$, with $ K_{a}=1-\gamma$. If we can do a measurement
which does not cause any additional scattering these values would yield
respectively a conductance in the bulk (of the junction) equal to $g_0K_2$
($ g_0K_{a}$). Thus we recover the bulk result by inverting our assumption
about $\delta $, not the temperature. But can we identify the two
quantities? An adiabatic switching of the external field is necessary if the
system has no way to relax. But any real system is coupled to its environment. A simple way
to describe this coupling is through a relaxation time $\tau _p$. It
is tempting to substitute $\delta \rightarrow \tau _p^{-1}$: now $\tau
_p^{-1}$ is not necessarily less than $\omega$ nor $\omega_2$. This depends on the underlying
mechanism, which does not yield simply an imaginary part added to the external
frequency $\omega$.
 We make a final remark which will be useful for the sequel: the instantaneous correlation functions vanish
exponentially at separations greater than the thermal length
$L_T=u/T$. Thus $L_T$ plays the role of a coherence length.

\section{Andreev reflection}
At times $t$ much less than $t_2$, the neighborhood of the origin and of the
contacts behave differently. In particular, when $K_2>1$ and $
K_1=1$, the local pairing correlation function decreases from $t^{-2/K_2}$ on
the bulk to $t^{-2/K}$ at the contacts, where $K=(1+K_2)/2$. Since $K>1$, this indicates a tendency towards superconducting order which
extends from the wire to the external leads. On the other hand, an incident
electron on $-a$ is reflected with a coefficient $\gamma $ which is now
negative, thus a partial hole is reflected back. These two facts are
respectively the analogous of proximity effect and Andreev reflection\cite{andreev}: it is
known that an electron incident on a normal metal--superconductor interface
needs to make a pair with an electron to enter the superconductor. Depending
on whether its energy is less or greater than the gap, a total or a partial
hole is reflected  \cite{andreev}. In our case, there is no gap, so we only
get a partial hole reflected. However, in the limit $K_2\rightarrow
\infty $, we get exactly one hole reflected.

We can also consider the case where the central wire is finite but
noninteracting ($K_2=1$), while the external leads have $K_1>1$: this is
reminiscent of an S-N-S structure  \cite{gun_sns}. The internal reflections we
found can be thought of as multiple Andreev reflections, and enhance the
conductance of the wire: $g=K_1g_0$.

To summarize, the physics is controlled by the local (external) $K$ at time
much less (larger) than the traversal time $t_2$ of the central wire. An
incident electron from an external lead is partially reflected at the
contacts to accommodate the interactions with the electrons inside the
central wire. But there is no reason to expect only the change in
interactions to cause scattering: any mismatch at the contact with the leads
(e.g.  geometrical) can reflect an incident electron. Impurities inside the
wire have also to be considered.

\section{Effects of a backscattering potential}
 The first question we address in this section is the validity of
Landauer formula in the presence of disorder. Next, the conductance is
computed in a perturbative way for a Gaussian random disorder, as well as
for a barrier at one contact. The results of the latter case are confirmed
by a renormalization group approach, explicitly done at finite temperature.

\subsection{Landauer formula}

In the presence of either local or extended disorder on the central wire, we
add to $H$, eq.(\ref{H}), the term 
\begin{equation}
H_b=\int dxV(x)\Psi ^{\dagger }(x)\Psi (x).  \label{Hb}
\end{equation}
The equation of motion for $\Phi $ (\ref{motion}) becomes nonlinear, with an
additional force 
\begin{equation}
F(x)=-\frac{\partial H_b}{\partial \Phi (x)}  \label{force}
\end{equation}
We express the new Green function in terms of $G(x,y,t)$ and $F(x,t)$. The
measurement procedure is the same as in the pure case. We skip the details,
and give the exact expression for the conductance: 
\begin{equation}
g/g_0=1-\int \int dxdy\frac d{d\omega }\langle \langle F(x)F(y)\rangle
\rangle _{\omega =0}  \label{exactF}
\end{equation}
We have used the method of G\"otze and W\"olfle \cite{gotze}, and have
 extended Apel's procedure \cite{apel} to deal with our inhomogeneous
 system. The force-force correlation in eq.(\ref{exactF}) is proportional to
 the $2k_f$ density response function. Following Apel, we interpret the
 integral appearing in $g$ as a reflection coefficient $R$ due to the
 impurities. Than eq.(\ref{exactF}) says that $g=g_0\left( 1-R\right)
 =g_0T$, which is the Landauer identity (\ref{land}) now extended to an
 interacting dirty wire. Note that if we impose the current and measure the
 potential drop as in \cite{apel}, we find instead $g=g_0\left( 1-R\right)
 /R$. Of course, $R$ is different from what we obtain in the absence of
 leads. Only a perturbative computation of $R$ will be performed.

\subsection{Perturbative correction to the conductance}
In an homogeneous interacting wire with parameters $u$ and $K$, the
sensitivity to a backscattering potential is determined by $K$.  Extended
and local disorder renormalize to zero or infinity depending on whether $K$
is greater or smaller than $3/2$ and $1$, respectively
\cite{giamarchi_loc,apel_rice,kane_fisher}. Since the wire length $L$ is the
largest length in the problem, the renormalization has a lower energy cutoff
$uL^{-1}$. A widespread recipe to deduce the behavior at length $L$ from the
properties in the thermodynamic limit consists in replacing $T\to uL^{-1}$
\cite{ogata,kane_fisher}. It is thus an interesting issue to see if our
model of a finite wire can yield a confirmation of this approach.

\subsubsection{Gaussian disorder}

We consider now a disordered wire, where backscattering impurities are
randomly distributed on the segment $[-a,a]$. For simplicity and to compare
with the known results we suppose that the potential in (\ref{Hb}) has a
uniform Gaussian distribution
\[
\langle V(x)V(y)\rangle =D\delta (x-y)
\]
The exact expression in (\ref{exactF}) has to be averaged. To first order in
$D$, the $2k_f$ density response function can be computed with the pure
Hamiltonian $H$ and is thus related to $G$. But the resulting expression is
so complicated that we can evaluate $g$ only in the high and low temperature
limits. Even then the calculation is involved, because one has to integrate
over time and space. 

At $T\gg \omega _2$ we find that $R$ contains the term found by Apel and
Rice \cite{apel_rice}, $\approx 2aT^{2(K_2-1)}$, but contains in addition
infinitely many powers of $T$. If we keep only the three main terms, we have
\begin{equation}
R=D\left[c_1 \frac{T}{\omega_2}(\tau_0T)^{2K_2-3}+c_2(\tau_0T)^{2(K_{a}-1)}+
c_3(\tau_0 T)^{2K_{2}-3}\right]   \label{dis}
\end{equation}
$\tau_0$ is a time cutoff of the order of the inverse bandwidth and
therefore much smaller than $\beta$ and $t_2$. The $c_i$ are constants
depending only on $K_2$. For convenience, we call the above three terms
$R_i$. We have always $R_3\ll R_1$. If $K_2<(>)K_c=(3+\sqrt{17})/4$, then
$R_2\ll(\gg)R_3$. In particular, $R_1$ is the dominant term if
$K_2<K_c$. But if $K_2>K_c$, the ratio $R_1/R_2$ depends explicitly on $T$ and $\omega_2$, and we have to consider $R=R_1+R_2$. 
For $T\ll \omega _2$ the correction is independent of temperature because of
the noninteracting leads, but depends on the length in the following way:
\begin{equation}
R=D_\xi [c\left(\tau_0\omega _2\right)^{2K_2-3}+c^{\prime }\left(\tau_0\omega _2\right)^{2(K_{a}-1)}]  \label{disw}
\end{equation}
The first (second) power is the dominant term if $K_2$ is less (greater)
than $K_c$.  We see that the substitution $T\to\omega _2$ allows
one to go from (\ref{disw}) to (\ref{dis}).

\subsubsection{A weak barrier at -a}

The simplest way to model an imperfection at a contact is through a point
like backscattering potential. Thus $V(x)$ in eq.(\ref{Hb}) is now centered
around $-a$. In a spirit similar to Kane and Fisher \cite{kane_fisher}, we
develop the boson field $\Phi $ around $-a$ (separately on the left and
right side of $-a$ since $\partial _x\Phi $ is discontinuous). Then
\[
H_b=\sum_{m=-\infty }^{+\infty }\frac{V_m}{\pi\tau_0} \exp 2im\Phi (-a) 
\]
where we include the higher harmonics.

If we evaluate eq.(\ref{exactF}) to first order in $V_m$, we get, for $T\gg
\omega _2$
\begin{equation}
g/g_0=1-\sum_m m^2V_m^2B\left(\frac 12,m^2K_{a}\right)\left( \pi \alpha T\right)
^{2(m^2K_{a}-1)}  \label{gimp}
\end{equation}
$B$ is the beta function. The correction is the exactly the same as that for a
 homogeneous wire with $K_{a}$ in the presence of a barrier. Note the
 independence of $g$ on the wire length.

For $T\ll \omega _2$ we have 
\begin{equation}
g/g_0=1-\sum_mm^2V_m^2c_mB\left(\frac 12,m^2\right)\left( \pi \tau _0T\right)
^{2(m^2-1)}\left( \tau _0\omega _2\right) ^{2(m^2K_{a}-1)}  \label{gimpw}
\end{equation}
where $c_m$ depends only on $K_2$. As in the previous section, the
substitution $T\to \omega _2$ works well for
the first harmonic contribution, but not for the higher ones.
\begin{figure}[htb]
\begin{center}

\mbox{\epsfig{file=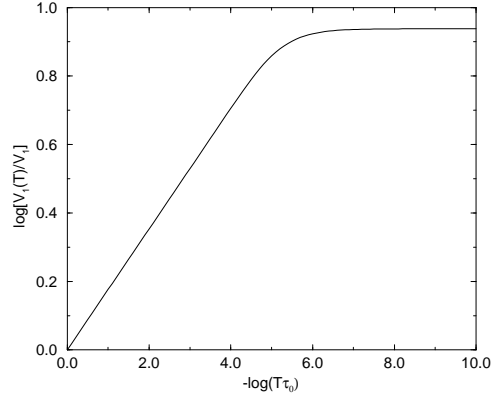,height=6cm}}

\caption{A weak barrier at $-a$:
$V_1(\beta/2)$ for $K_1=1$, $K_2=0.7$ and $t_2=100\tau_0$ plotted in
logarithmic scale as function of the temperature. The crossover from
$(1-K_a)l$ to a plateaus occurs at
$l=log[\omega_2\tau_0]=4.6$.} \label{VT}
\end{center}
\end{figure}
We can also carry out the renormalization procedure using
imaginary time at finite temperature: we increase $\tau_0$ to $\tau_0 e^l$
and modify the parameters in order to keep the partition function
invariant. Only the coefficients $V_m$ need to be renormalized. If we neglect
higher-order terms, we find that the renormalization equations integrate to
\begin{equation}
V_m(l)=V_m(0)\exp \left[ l-m^2U(-a,-a,l)\right]  \label{Vm}
\end{equation}
where $U(x,x;l)=2[G(x,x,\tau_0 )-G(x,x,\tau_0 e^l)]$, with $G$ the imaginary
time Green function. When $K$ is uniform:

\begin{equation}
U(\tau )=K\log \left| \sin [\pi \tau T]/\pi \tau _0T\right|  \label{homU}
\end{equation}
In our model, $U$ is a series of logarithmic functions we avoid to write
down. The effective $V_1$ obtained from eq.(\ref{Vm}) when we stop the renormalization at
the maximum allowed cutoff value $\beta/2$ is plotted in fig.(\ref{VT}). It exhibits a crossover from a power
law dependence at $T> \omega_2$ to a ``marginal'' behavior at $T< \omega_2$.

 In order to obtain further information,
we approach $U$ in the extreme limits of low or high temperature compared to $\omega _2$.
Consider first $T\gg \omega _2$: the coherence between the two contacts is
lost, and the effective $K$ in (\ref{homU}) is local:
$K_{a}=2K_2/(1+K_2)$. If we stop the renormalization equation at $\beta /2$, eq.(\ref{Vm}) and
eq.(\ref{homU}) yield
\begin{equation}
V_m(\beta /2)=V_m\left( \pi T\tau _0\right) ^{(m^2K_{a}-1)}  \label{Vl}
\end{equation}
Thus the renormalized $V_m$ have the usual power-law behavior controlled by
the local conductance $K_{a}$.

In the opposite limit of low temperature, there is coherence on a scale
larger than the traversal time $t_2$. $U$ contains two parts: a temperature
dependent term similar to eq.(\ref{homU}) with the external $K_1=1$, and an
$\omega_2$ dependent term. Up to a constant depending only on $K_2$, the
renormalized $V_m$ turns out to be, in view of eq.(\ref{Vm}):
\begin{equation}
V_m(\beta /2)=V_m(\tau _0)(\pi \tau _0T)^{(m^2-1)}(\omega _2\tau
_0)^{(m^2K_{a}-1)}  \label{Vh}
\end{equation}
In particular, $V_1$ doesn't depend on the temperature, but acquires an $
\omega _2$ dependence due to the integration of the energies higher than $
\omega _2$.

From eqs.(\ref{Vl}) and (\ref{Vh}), it appears that it is $K_{a}$ {\em \
that controls the power law behavior} either with temperature or length, and
this is true for the first harmonic contribution only. The perturbation
approach is valid as long as $K_{a}>1$, but this condition is equivalent to
$ K_2>1$. If $K_2<1$, one has to control the smallness of $V_m$ in
eqs.(\ref{Vl}) and (\ref{Vh}). We recover the same correction to the
conductance obtained in eqs.(\ref{gimp} ) and (\ref{gimpw}) replacing the
$V_m$ by their renormalized values from eqs.  (\ref{Vl}) and (\ref{Vh}), but
computing the $2mk_f$ response functions at the cutoff $\beta /2$.

We note that we can equally treat a weak impurity localized at a point $x$
inside $\left[ -a,a\right] $.The renormalized $V_m$ has the form of
eq.(\ref{Vm}), with $U(x,x,l)$ instead of $U(-a,-a,l)$. The conductance
correction can be found, but there is no place to give more details here.

\subsection{Summary}

In contrast to the pure case where the conductance does not depend on the
interactions inside the wire, the backscattering potential generates a power
law behavior determined by $K_2$. In the presence of a weak backscattering
at one contact, the correction is similar to that predicted in a homogeneous
wire with a parameter $K_{a}$ and a barrier. If there is a Gaussian
distributed disorder, one encounters again a contact exponent $2(K_{a}-1)$
in addition to the bulk term usually found. At $K_2<K_c$, the latter
dominates the former. This concerns in particular repulsive interactions,
which seems to be the case in a quantum wire. But at $K_2>K_c$ the boundary
of the wire has a non-negligible contribution. From eq.(\ref{Vl}), the Andreev reflection
we talked about persists to a weak barrier at the contact because
$K_{a}>1$. The computation can be carried out similarly for interacting
leads: this might have an interest in an S-N-S structure. Note that when
$K_1\neq 1$, we cannot go from the high to the low temperature regime by
the simple substitution temperature $\to \omega_2$.

\section{Summary}

We have investigated transport through a finite interacting wire connected
to noninteracting leads. The conductance of the pure wire is not
renormalized by the interactions for any spatial variation of the internal
parameters $u,K$. If $K$ varies abruptly, the correlation parameter $K_2$
of the wire controls the decrease of the conductance in the presence of a
backscattering potential. In quantum wires where interactions are believed
to be repulsive, we recover the usual power law behavior. This seems to fit
well the experiments of Tarucha \cite{tarucha}. The agreement of our
results with experiment is surprising in view of our crude treatment of the
opening up of the measuring leads into a two-dimensional electron gas.

Other authors \cite{maslov_g,ponomarenko} adopted the same model and found
the same result for the conductance of the pure wire. Nevertheless, we do not
agree with the conductance correction as derived by Maslov in
\cite{maslov_disorder}: the inhomogeneity of the correlation functions is
ignored and only the bulk contribution is found.

\bibliography{revues,data}
\bibliographystyle{prsty}

\end{document}